\documentclass[12pt,preprint]{aastex}
 \usepackage{emulateapj5}
  \usepackage{graphicx}
\usepackage{longtable}
\usepackage{ulem}
\usepackage{float}
\usepackage{color}
\newcommand{\be}{\begin{equation}}
\newcommand{\ee}{\end{equation}}
\def\ergs{{\rm\,erg\,s^{-1}}}
\newcommand{\msun}{{M}_{\sun}}

\slugcomment{Printed at \today}
\shorttitle{Accretion-jet model in M87} \shortauthors{Feng et al.}

\begin{document}

\title{An accretion-jet model for M87: interpreting the spectral energy distribution and Faraday rotation measure}

\author{Jianchao Feng\altaffilmark{1}, Qingwen Wu\altaffilmark{1}, and Ru-Sen Lu\altaffilmark{2}}
\altaffiltext{1}{School of Physics, Huazhong University of Science and Technology,
 Wuhan 430074, China}
\altaffiltext{2}{Max-Planck-Institut f\"ur Radioastronomie, Auf dem Hugel 69, 53121 Bonn, Germany}
\altaffiltext{3}{Corresponding author, email: qwwu@hust.edu.cn}

\begin{abstract}
M87 is arguably the best supermassive black hole (BH) to explore the jet and/or accretion physics due to its proximity and fruitful high-resolution multi-waveband observations. We model the multi-wavelength spectral energy distribution (SED) of M87 core that observed at a scale of 0.4 arcsec ($\sim 10^5R_{\rm g}$, $R_{\rm g}$ is gravitational radius) as recently presented by Prieto et al. Similar to Sgr A*, we find that the millimeter bump as observed by Atacama Large Millimeter/submillimeter Array (ALMA) can be modeled by the synchrotron emission of the thermal electrons in advection dominated accretion flow (ADAF), while the low-frequency radio emission and X-ray emission may dominantly come from the jet. The millimeter radiation from ADAF dominantly come from the region within $10R_{\rm g}$, which is roughly consistent with the recent very long baseline interferometry observations at 230\,GHz. We further calculate the Faraday rotation measure (RM) from both ADAF and jet models, and find that the RM predicted from the ADAF is roughly consistent with the measured value while the RM predicted from the jet is much higher if jet velocity close to the BH is low or moderate (e.g., $v_{\rm jet}\lesssim0.6\,c$). With the constraints from the SED modeling and RM, we find that the accretion rate close to the BH horizon is $\sim (0.2-1)\times10^{-3}\msun \rm yr^{-1}\ll\dot{\it M}_{\rm B}\sim 0.2\it \msun \rm yr^{-1}$ ($\dot{M}_{\rm B}$ is Bondi accretion rate), where the electron density profile, $n_{\rm e}\propto r^{\sim -1}$, in the accretion flow is consistent with that determined from X-ray observation inside the Bondi radius and recent numerical simulations.
\end{abstract}


\section{Introduction}
The giant radio galaxy M87 is one of the well-known radio loud low-luminosity active galactic nuclei (AGNs). It is an excellent laboratory for investigating the accretion and jet physics because of its proximity with a distance of $D=16.7\pm0.6$ Mpc \citep[]{jord05,blak09} and a large estimated black hole (BH) mass of $3-6.6\times10^9 \msun$ \citep{macc97,gebh11,wals13}. The bolometric luminosity of the core is estimated to be $L_{\rm bol}\sim2.7\times10^{42} \ergs \sim 3.6\times10^{-6} L_{\rm Edd}$ \citep[$L_{\rm Edd}$ is Eddington luminosity,][]{pri16}, which is several orders of magnitude less than those of Seyferts and quasars. The quite low Eddington ratio in M87 suggests that it most possibly accretes through a radiatively inefficient accretion flow \citep[see][for a recent review and references therein]{yn14}. Recent high spatial resolution $Chandra$ X-ray observations have resolved the Bondi radius, $R_{\rm Bondi} \approx$ 0.2\ kpc $\approx 8\times 10^{5} R_{\rm g}$, where $R_{\rm g}=GM_{\rm BH}/c^{2}$ is the gravitational radius \citep[]{russ15}. In combination with the inferred gas density at the Bondi radius as about 0.3 $\rm cm^{-3}$, the Bondi accretion rate is estimated to be $\dot{M}_{\rm B} \approx 0.2 M_\odot \rm yr^{-1}$ \citep[e.g.,][]{russ15}, which indicates that either the radiative efficiency of the accretion flow is very low ($\eta\sim L_{\rm bol}/\dot{M}_{\rm B}c^2 \approx 10^{-4}$) or most of the matter at Bondi radius is not captured by the BH, or both.

The Galactic center BH (Sgr A*) and the supermassive BH in the center of the Virgo cluster (M87), are the two largest BHs on the sky, with putative event horizons subtending $\sim$ 53 and 38 microarcseconds ($\mu$as) respectively \citep[e.g.,][]{rd15}. The Event Horizon Telescope (EHT), a planed Earth-sized array at millimeter (mm) and submillimeter (submm) wavebands, provides well-matched horizon-scale resolution for Sgr A* and M87 \citep[e.g.,][]{doe09}, which greatly help to study the accretion and/or jet physics in both sources \citep[e.g.,][]{doe08,hua09,mos09,dex10,fish11,bro11,lu14}. In particular, M87 is the best source for exploring the jet physics near a BH due to its strong jet has been observed in multiwaveband, where the multi-wavelength studies also have been made from radio to $\gamma$-ray~\citep[e.g.,][]{rei89,jun99,per05,har06,ly07,kov07,doe12,ak15,had16}. Recently, it is possible to explore the inner jet physics with high-resolution EHT observations at 230\,GHz, which resolve the jet base at $\sim 10 R_{\rm g}$ \citep[]{doe12,ak15}. \citet{asa12} investigated the structure of the M87 jet from milliarcsec (mas) to arcsec scales by utilizing multi-frequency very long baseline interferometry (VLBI) images, where they found that the jet follows a parabolic shape, $Z\propto R_{\rm j}^{1.73\pm0.05}$, in a deprojected distance of $\sim 10^2-10^5 R_{\rm g}$ ($R_{\rm j}$ is the radius of the jet emission and $Z$ is the axial distance from the core). The acceleration zone of M87 jet may be co-spatial with the jet parabolic region, where the intrinsic jet velocity fields increase from $\sim 0.1\ c$ at $\sim 10^2R_{\rm g}$ to 1 $c$ at $\sim 10^5 R_{\rm g}$ \citep[]{asa14}. The multi-wavelength nuclear SED of M87 has been extensively explored by both pure jet models \citep[e.g.,][]{dex12,de15,pri16} and ADAF+jet models \citep[e.g.,][]{dim03,yua09,br09,li09,nem14,moc16}, where the radio emission is produced by the jet in both models while the millimeter/sub-millimeter and X-ray emission can either come from the jet or ADAF.

Apart from the continuum spectrum, linear polarization can be a diagnostic of the relativistic jets and accretion flows associated with BH systems. In particular, millimeter/submillimeter polarimetry provides an important tool to study the magnetized plasma near a BH through the Faraday rotation of the polarized light. It was found that the accretion rate close the BH ($\lesssim 10 R_{\rm g}$) is several orders of magnitude lower than the accretion rate at Bondi radius ($R_{\rm B}\sim 10^{5-6} R_{\rm g}$) in Sgr A* based on the Faraday rotation measure stuidies \citep[RM, an integral of the product of the thermal electron density and the magnetic field component along the line of sight,][]{bow03,mac06,mar06}. \citet{kuo14} presented the first constraint on the Faraday RM at millimeter wavelength for the nucleus of M87 and found that the best fit RM is $-(2.1\pm1.8)\times10^5\rm rad\ m^{-2}$ (1$\sigma$ uncertainty). Using the same method as in Sgr A* \citep[][]{mar06}, \citet{kuo14} found the accretion rate should be below $9.2\times10^{-4}\msun \rm yr^{-1}$ at a distance of 21 Schwarzschild radii from the BH, which suggest that most of the matter at Bondi radius is not really accreted by the BH.

\begin{figure*}
\centering
\includegraphics[width=90mm]{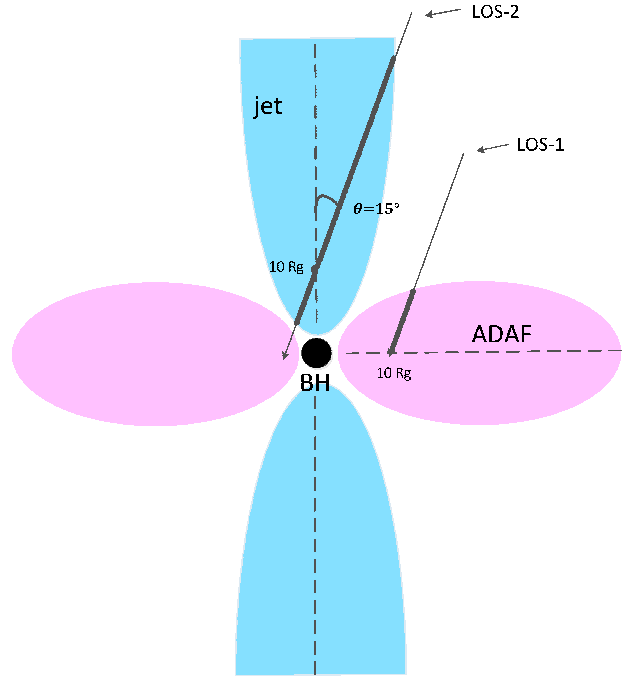}
\caption{A cartoon picture of our ADAF-jet model, where a geometrical thick, optically thin ADAF and parabolic shape of jet are considered. The jet inclination angle is assumed to be $15^{\rm o}$, and the disk is perpendicular to the jet. Here, we consider the two possibilities where the polarized emission pass though ADAF itself along LOS-1 and the polarized emission of ADAF pass through the plasma in the jet along LOS-2 (the thick solid lines).}
\end{figure*}

Recently, \citet{pri16} presented the high-resolution quasi-simultaneous multi-waveband SED at scale of $\sim 0.4$  arcsec for M87, which is very helpful to explore the accretion-jet physics. In particular, the evident millimeter bump in the SED of M87 is quite similar to the sub-millimeter bump of Sgr A* \citep[e.g.,][]{yu03}, which may be contributed by the synchrotron emission from the thermal electrons in ADAF. If this is the case, it can be used to constrain the accretion rate near the BH, since that most of former works believed that the multi-waveband emission of M87 core is dominated by the jet which prevent us to learn about the underlying accretion physics. Furthermore, the recently reported Faraday rotation measure will put another constraint on the accretion and jet model. We present the ADAF-jet model in Section 2, and show the main results in Section 3. Discussion and conclusion will be given in Section 4. Throughout this work, we adopt a BH mass of $6.6\times10^9 \msun$ and a distance of 16.7 Mpc, where 1 mas = 0.08 pc = 280 $R_{\rm g}$.

\begin{table*}[t]
\centering
\begin{minipage}{180mm}
\centering
\footnotesize
  \centerline{\bf Table 1. M87 core SED in quiescent phase with aperture radius of $\sim 0.4^{''}$.}
\tabcolsep 1.500mm
\begin{tabular}{llllc}\hline\hline
\tablecolumns{16}
 Frequency       &Flux                         & Telescope & Date          &References\\
\hline
$5.0\times10^{9}$Hz      &$3.10\pm0.06 $            Jy    &VLA-A   &  1999-09         & 1         \\
$8.4\times10^{9}$Hz     &$3.02\pm0.02 $             Jy   &VLA-A   &  2003-06\&2003-08 & 2          \\
$8.4\times10^{9}$Hz     &$3.15\pm0.16 $             Jy   &VLA-A   &  2004-12-31       & 2          \\
$15.0\times10^{9}$Hz    &$2.7\pm0.1   $             Jy   &VLA-A   &  2003-06\&2003-08 & 2          \\
$22.0\times10^{9}$Hz    &$2.0\pm0.1   $             Jy   &VLA-A   &  2003-06        & 2          \\
$93.7\times10^{9}$Hz    &$1.82\pm0.06 $             Jy   &ALMA    &  2012-6-3       & 2          \\
$108.0\times10^{9}$Hz   &$1.91\pm 0.05$             Jy   &ALMA    &  2012-6-3       & 2          \\
$221.0\times10^{9}$Hz   &$1.63\pm0.03 $             Jy   &ALMA    &  2012-6-3       & 2           \\
$252.0\times10^{9}$Hz   &$1.42\pm0.02 $             Jy   &ALMA    &  2012-6-3       & 2           \\
$286.0\times10^{9}$Hz   &$1.28\pm0.02 $             Jy   &ALMA    &  2012-6-3       & 2          \\
$350.0\times10^{9}$Hz   &$0.96\pm0.02 $             Jy   &ALMA    &  2012-6-3       & 2           \\
$635.0\times10^{9}$Hz   &$0.43\pm0.09 $             Jy   &ALMA    &  2012-6-3       & 2          \\
$2.6\times10^{13}$Hz    &$(1.3\pm0.2)\times10^{-2}$   Jy   &Keck    &  2000-01-18     & 3          \\
$2.8\times10^{13}$Hz    &$(1.67\pm0.09)\times10^{-2}$ Jy   &Gemini  &  2001-05        & 4           \\
$1.37\times10^{14}$Hz   &$(3.3\pm0.6)\times10^{-3}$   Jy   &HST     &  1998-1-16      & 2           \\
$1.81\times10^{14}$Hz   &$(3.1\pm0.8)\times10^{-3}$   Jy   &HST     &  1999-1-16      & 2           \\
$2.47\times10^{14}$Hz   &$(2.06\pm0.18)\times10^{-3}$ Jy   &HST     &  1997-11-10     & 2           \\
$3.32\times10^{14}$Hz   &$(1.38\pm0.01)\times10^{-3}$ Jy   &HST     &  2003-1-19      & 2           \\
$3.70\times10^{14}$Hz   &$(9.5\pm1.9)\times10^{-4}$   Jy   &HST     &  2003-11-29     & 2           \\
$4.99\times10^{14}$Hz   &$(6.33\pm0.63)\times10^{-4}$ Jy   &HST     &  2003-11-29     & 2           \\
$6.32\times10^{14}$Hz   &$(4.13\pm0.12)\times10^{-4}$ Jy   &HST     &  2003-11-29     & 2           \\
$8.93\times10^{14}$Hz   &$(2.10\pm0.04)\times10^{-4}$ Jy   &HST     &  2003-5-10      & 2           \\
$8.93\times10^{14}$Hz   &$(2.16\pm0.04)\times10^{-4}$ Jy   &HST     &  2003-3-31      & 2           \\
$1.11\times10^{15}$Hz   &$(1.55\pm0.03)\times10^{-4}$  Jy  &HST     &  2003-05-10     & 2           \\
$1.27\times10^{15}$Hz   &$(1.05\pm0.03)\times10^{-4}$  Jy  &HST     &  2003-7-27      & 2           \\
$1.36\times10^{15}$Hz   &$(1.33\pm0.04)\times10^{-4}$ Jy   &HST     &  2003-11-29     & 2           \\
$2.06\times10^{15}$Hz   &$(4.73\pm0.47)\times10^{-5}$ Jy   &HST     &  1999-5-17      & 2           \\
  2-10 keV            & $(0.70\pm0.04)\times10^{-12}\rm erg\ cm^{-2}\ s^{-1}$ & Chandra & 2000-07-30     & 5           \\

\hline
\end{tabular}
\end{minipage}

\begin{minipage}{170mm}
\centering
 References: 1) \citet{naga01}; 2) \citet{pri16}; 3) \citet{whys04}; 4) \citet{per01}; 5) \citep{russ15}.
\end{minipage}

\end{table*}

\begin{figure*}
\epsscale{1.0} \plotone{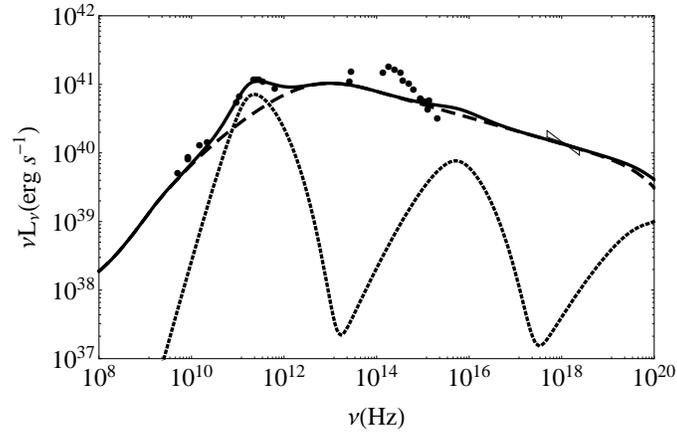}
 \caption{ADAF-jet model result compared with M87 0.4 arcsec aperture radius SEDs in quiescent state. The dotted-line represent the ADAF spectrum with $a_{*}=0.9$, $s=0.52$ and $\beta=0.5$. The dashed lines show the jet spectrum with $v_{\rm jet}=0.6\ c$, $\dot{m}_{\rm jet}=1.5\times10^{-6}$ and $p=2.38$.  The solid line is the sum of ADAF and jet contribution.}
\end{figure*}

\section{Accretion-jet model}
Due to the low Eddington ratios of M87, we adopt the ADAF model that is widely used in modeling the SED of the quiescent and low-luminosity AGNs \citep[e.g.,][]{ic77,ab95,ny95,yu03,wu06,wu07,ho08,wu13,lw13,cao14}. The global structure and dynamics of the accretion flow in general relativistic frame is calculated numerically to obtain the ion and electron temperature, density at each radius, since that the BH may be fast rotating in M87. The accretion rate at each radius is $\dot{M}=\dot{M}_{\rm out}(R/R_{\rm out})^{\it s}$, where $\dot{M}_{\rm out}$ is the accretion rate at the outer radius, $R_{\rm out}$, of the ADAF  and $s$ is the wind parameter. In this work, we simply set $R_{\rm out}=R_{\rm B}$ and $\dot{M}_{\rm out}=\dot{M}_{\rm B}$. The global structure of the ADAF can be calculated if the parameters $\alpha$, $\beta$, and $\delta$ are specified, where $\alpha$ is viscosity parameter, $\beta$ is the ratio of gas to total pressure (sum of gas and magnetic pressure), and $\delta$ describes the fraction of the turbulent dissipation that directly heats the electrons in the flow \citep*[see][for more details]{man00}. For $\alpha$, we adopt typical values of 0.3 as widely used in ADAF models. The value of $\beta$ is typically $\sim0.5-0.9$ \citep[][]{yn14}, where $\beta=0.5$ correspond to the equipartition between magnetic energy and thermal energy. Similar to modeling of Sgr A*, we adopt $\delta\sim0.3$ \citep[e.g.,][]{yu06}, which is roughly consistent with the simulations by \citet[][]{sh07}. We keep $s$ as a free parameter, which can be constrained in SED fitting if other parameters are fixed. We take into account three processes of the radiative cooling, i.e., the synchrotron radiation, the bremsstrahlung, and the multi-Comptonization of soft photons, where the general description of cooling processes and relevant formulae have been presented by \citet{ny95} and \citet{man00} in a more handy way. For caculation of the Comptonization, we adopt the program given by \citet{cb90}. In spectral calculations, the effect of the bending of light and the gravitational and the Doppler shift of the energy of the photons should be considered. In this work, we consider the gravitational and the Doppler shift of the energy of the photons, while the effect of the bending of light was neglected for simplicity \citep[see more details in][]{man00}, which does not affect our main conclusion.

The mechanisms of the jet formation, collimation, acceleration and dissipation are very unclear. In this work, we adopt a phenomenological jet model due to above uncertainties. We assume a small fraction of the material in the ADAF is transferred into the vertical direction to form a jet, since the velocity of the accretion flow is supersonic near the black hole and a standing shock should occur at the bottom of the jet because of bending. From the shock-jump conditions, we calculate the properties of the post shock flow, such as the electron temperature $T_{\rm e}$ \citep*[e.g.,][]{yc05}. With high-resolution VLBI observations, \citet{asa12} found that the collimation profile of the M87 jet is parabolic on scales up to $\sim 5 \times 10^{5} R_{\rm g}$($Z\propto R_{\rm j}^{1.73\pm0.05}$) and then transits to a conical shape beyond that. We adopt this observational parabolic shape in our model. For the jet radiation, we mainly adopt the internal shock scenario, which is widely used in interpreting gamma-ray burst (GRB) afterglows \citep*[e.g.,][]{pi99,sp01}, the multi-wavelength SED of XRBs \citep*[e.g.,][]{xy16} and AGNs \citep*[e.g.,][]{wu07}. The internal shock scenario is that the central power engine produces energy that is channelled into jets in an intermittent way, where the faster shells will catch up with slower ones, and internal shocks are formed in the jet at a scale of $\sim \Gamma^2 R_{\rm g}$ ($\Gamma$ is Lorentz factor). These shocks accelerate a fraction of the electrons, $\xi_{\rm e}$, into a power-law energy distribution with an index $p$. The radiative cooling is also considered self-consistently for the distribution of the accelerated electrons, where the power-law electrons should be truncated at higher energies due to the cooling. In this work, we adopt a typical value of $\xi_{\rm e}=0.01$ and allow the $p$ to be a free parameter that can be constrained from observations \citep*[see][for more details and references therein]{yc05}. The energy density of accelerated electrons and amplified magnetic field are determined by two parameters, $\epsilon_{\rm e}$ and $\epsilon_{\rm B}$, which describe the fraction of the shock energy that goes into electrons and magnetic fields, respectively. Obviously, $\epsilon_{\rm e}$ and $\xi_{\rm e}$ are not independent. In calculation of the jet spectrum, the emission and absorption of both the thermal electrons and nonthermal electrons are considered. It should be noted that only synchrotron emission is included in calculation of the jet spectrum, where the synchrotron self-Compton in the jet is several orders of magnitude less than the synchrotron emission in X-ray band \citep*[see,][for more discussions]{wu07}. The jet inclination angle of $\sim15^{\rm o}$ is adopted \citep*[e.g.,][]{wz09}. We treat the mass-loss rate, $\dot{m}_{\rm jet}=\dot{M}_{\rm jet}/\dot{M}_{\rm Edd}$, and jet velocity, $v_{\rm jet}$, as free parameters. In Figure 1, we show a cartoon picture of our model.

\begin{table*}[t]
\centering
\begin{minipage}{180mm}
\centering
\footnotesize
  \centerline{\bf Table 2. Model results from ADAF.}
\tabcolsep 1.500mm
\begin{tabular}{lcccc}\hline\hline
\tablecolumns{16}
\tabletypesize{\fontsize{0.4in}}
$a_{*}$ & $\rm \beta$& $s$ & $\dot{M}(\rm 10Rg)$ $(M_\odot\ \rm yr^{-1})$ & RM ($\rm rad/m^{2}$)\\
\hline
0.9   &0.5   &0.52 & $5.8\times10^{-4}$ &$2.3\times10^{5}$\\
0.9   &0.9   &0.48 & $9.0\times10^{-4}$ &$7.5\times10^{4}$\\
0     &0.5   &0.40 & $2.2\times10^{-3}$ &$1.5\times10^{6}$\\
0     &0.9   &0.37 & $3.1\times10^{-3}$ &$4.0\times10^{5}$\\

\hline
\end{tabular}
\end{minipage}

\begin{minipage}{170mm}
\centering
 Note: The RMs are calculated from the radius of $\rm 10Rg$ in the accretion flow from the line of sight .
\end{minipage}

\end{table*}

\begin{table*}[t]
\centering
\begin{minipage}{180mm}
\centering
\footnotesize
  \centerline{\bf Table 3. Model results from jet.}
\tabcolsep 1.500mm
\begin{tabular}{cccc}\hline\hline
\tablecolumns{16}
\tabletypesize{\fontsize{0.4in}}
$v_{\rm jet}$ ($c$) &   $\dot{M}_{\rm jet} (\rm M_\odot yr^{-1})$& $P_{\rm jet} (\rm erg/s)$& RM ($\rm rad/m^{2}$)\\
\hline

0.3  & $1.5\times10^{-3}$ &$4.4\times10^{42}$ & $1.6\times10^{9}$\\
0.6  & $2.2\times10^{-4}$ &$4.1\times10^{42}$ & $5.4\times10^{7}$\\
0.9  & $1.5\times10^{-5}$ &$2.6\times10^{42}$ & $3.6\times10^{6}$\\
0.99 & $1.5\times10^{-6}$ &$3.8\times10^{42}$ & $7.5\times10^{5}$\\
\hline
\end{tabular}
\end{minipage}

\begin{minipage}{170mm}
\centering
 Note: The RMs are calculated from the height of $\rm 10Rg$ in the jet along the line of sight.
\end{minipage}

\end{table*}

\section{Results}
\subsection{Modeling the multi-wavelength SED}
  In building the SED of M87 core, we adopt the multi-waveband data mainly from \cite{pri16}, where they presented the data at a scale of $\sim 0.4$ arcsec ($\sim$32 pc) for both quiescent and active states across the electromagnetic spectrum. In this work, we only focus on the data in quiescent phase due to the unclear physics for triggering the flares.
  Instead of using the average X-ray data, we adopt the lowest flux from $Chandra$ (30/07/2000) as the radiation in the quiescent state \citep[][]{russ15}. We list all the data in Table 1. The main features of the spectrum include: a flat radio spectrum ($S_{\nu}\propto \nu^{-\kappa}$, $\kappa\sim$0.2), the spectrum become steeper at $\sim100$ GHz ($\kappa\sim$-0.3) and turn over in millimeter region, a steep power-law spectrum from IR to the UV ($\kappa\sim1.6$) and a steep X-ray spectrum ($\kappa\sim1.4$).

   The solid line in Figure 2 shows the total spectrum corresponding to the ADAF-jet model of M87, where the long-dashed line represents the ADAF spectrum with $a_{*}=0.9$, $\beta=0.5$, and $s=0.52$, while the long-dashed line represents the jet spectrum with a moderate velocity of  $v_{\rm jet}=0.6\ c$ and $\dot{m}_{\rm jet}=3\times10^{-6}$.
   The millimeter bump can be naturally modeled by the synchrotron emission from thermal electrons in the ADAF. We find that the different parameters of $a_{*}=0-0.9$ and $\beta=0.5-0.9$ in the ADAF will lead to similar SED fitting if changing the parameter $s$ (or accretion rate at inner region, see Table 2), where the ADAF mainly contributes at millimeter waveband but little at other wavebands. The synchrotron radiation from the jet accounts well for the radio and X-ray emission. The jet spectrum is not sensitive to the jet velocity, as it is roughly unchanged for different jet velocities if we adjust the outflow rate simultaneously (e.g., $\dot{m}_{\rm jet}\sim10^{-7}-10^{-5}$ for $v_{\rm jet}=0.3-0.9\ c$, see Table 3). The steep IR to optical data cannot be well reproduced with our ADAF-jet model.


   In Figure 3, we present the 230\,GHz intensity distribution from the ADAF and jet respectively with the parameters obtained from above SED modeling. We find that the 230\,GHz emission in the ADAF dominantly comes from the region within 10$R_{\rm g}$ (top panel). However, the 230\,GHz emission in the jet comes mainly from the region much larger than $10R_{\rm g}$ (bottom panel). \citet{doe12} found that the 230\,GHz emission dominantly come from a very compact region within $\sim 10R_{\rm g}$, which prefer that the millimeter emission come from the thermal electrons in ADAF not the synchrotron emission in the jet of our ADAF-jet model.

\subsection{Constraints from Faraday rotation measure}

The possible contribution to the observed RM include both the ADAF surrounding around the BH and the jet that possibly perpendicular to the disk. The RM has been used to constrain the accretion rate in ADAF or outflow rate in jet for Sgr A* and M87 respectively \citep[e.g.,][]{yu03,kuo14,li15}. The RM depends on the distribution of the electron density and magnetic field, which is
  \be
   RM=8.1 \times 10^{5} \int \frac{\log \gamma_{e}(z)}{2\gamma^{2}_{e}(z)} n_{e}(z) B_{\parallel} dl\ \mathrm{rad}\ \mathrm{m}^{-2},
   \ee
 where $\gamma_{\rm e}=\kappa T_{\rm e}/m_{\rm e}c^2$ is electron Lorentz factor, $n_{\rm e}$ is the electron density in unit of $\rm cm^{-3}$, the path length $dl$ in unit of pc, and the magnetic field along LOS $B_{\parallel}$ in unit of Gauss. The factor $\log \gamma_{\rm e}(z)/\gamma^{2}_{\rm e}(z)$ is the relativistic correction \citep[][]{quat00,hua11}.

  Our SED modeling suggest that the mm emission mainly originate from the inner region of ADAF(e.g., within several $R_{\rm g}$), which is roughly consistent with recent observations \citep[e.g.,][]{doe12}. The RM that provided in \citet{kuo14} is also inferred from the polarization observation in this waveband. Therefore, the RM may be mainly contributed by the hot plasma in the inner region of the ADAF. Instead of assuming the spherical accretion flow in \citet{kuo14}, we calculate the RM from our ADAF along the line of sight (LOS, see LOS-1 in the cartoon of Figure 1), where the ADAF is a thick disk ($H/R<1$). In calculation of RM, we need know the distribution of electron density and magnetic field. Here, we simply assume the $B_{\parallel}\simeq B_{\rm ADAF}$ since a large-scale magnetic field is normally needed in the formation of a collimated, relativistic jet. The real RM should be a little bit lower due to the inclination angle between the magnetic field line and the LOS-1. The observational value of RM is $\sim-2.1\pm1.8\times10^5  \rm rad\ m^{-2}$, which was derived at $\sim$ 230\,GHz \citep[][]{kuo14}. In this work, we calculate the RM along the LOS-1 from $R=10R_{\rm g}$ (see Figure 1), where most of the millimeter emission originates from a compact region \citep[][]{doe12}. For the case of $a_{*}=0.9$, $\rm RM=3.3\times10^5\ \rm rad\ m^{-2}$ and $7.5\times10^4\ \rm rad\ m^{-2}$ for $\beta=0.5$ and 0.9 respectively. For a non-rotating BH with $a_{*}=0$, $\rm RM=1.5\times10^6$ and $4.0\times10^5\rm\  rad\ m^{-2}$ for $\beta=0.5$ and 0.9 respectively, of which a little bit higher accretion rates (or weaker wind) near the BH are needed to reproduce the millimeter bump in the SED when compared to the case of $a_{*}=0.9$ (see Table 2). It should be noted that the RM may be decreased by a factor of 2 if the LOS-1 is along a smaller radius (e.g., $R=2 R_{\rm g}$), which is suppressed by the relativistic effect due to higher electron temperature (see equation 1). Our main results are unchanged.

\begin{figure*}
\epsscale{1.0} \plotone{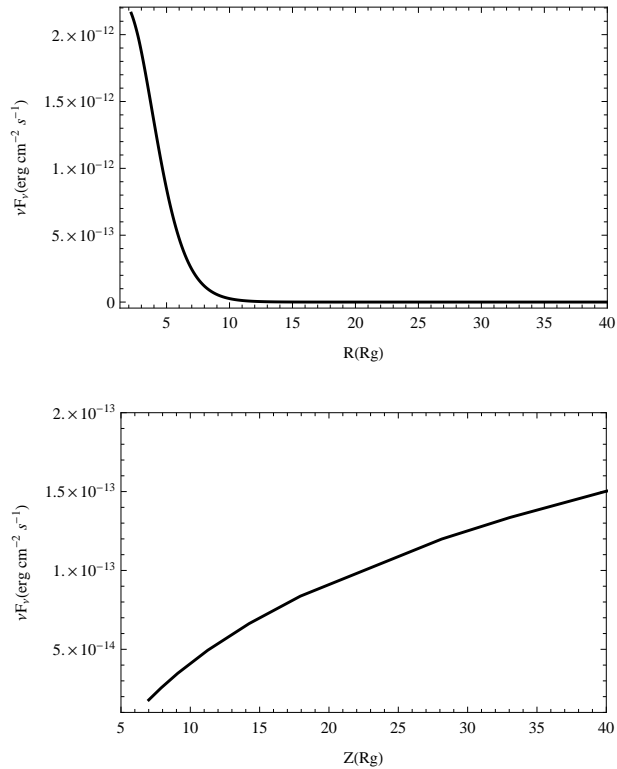}
 \caption{The intensity distribution at 230 GHz from ADAF (top panel) and jet (bottom panel) model respectively, where the model parameters are the same as those adopted in SED modeling in Figure 2.}
\end{figure*}

The other possible contribution to the observed RM, in additional to the ADAF, may come from the jet. Due to the jet emission at 230 GHz is much larger than the observational size \citep[see bottom panel of Figure 3 and][]{doe12}, we consider the possibility of the jet as an external origin of the RM, where the polarized emission in the disk pass through the jet along LOS-2 (see cartoon in Figure 1). We calculate the RM along LOS-2, where the LOS-2 intersects with the jet axis at the point of $Z\sim10 R_{\rm g}$ in jet axis due to the millimeter observations is quite compact \citep[e.g., within several $R_{\rm g}$,][]{doe12}.  In the jet model, we also assume $B_{\parallel}\sim B_{\rm jet}$ if poloidal magnetic field dominates near the BH horizon. The $RM=1.6\times10^9\rm rad\ m^{-2}$, $5.4\times10^7\rm rad\ m^{-2}$, $3.6\times10^6\rm rad\ m^{-2}$ and $7.5\times10^5\rm rad\ m^{-2}$ for the jet velocity of $v_{\rm jet}=$0.3 $c$, 0.6 $c$, 0.9 $c$ and 0.99 $c$ respectively (see Table 3). The RM becomes lower when jet velocity increases, which is caused by the lower outflow rate is needed in fitting the SED for the higher jet velocity (Doppler boosting effect).

\section{Discussion}

   The multi-wavelength SED of M87 has been widely modeled in literatures by ADAF model \citep[][]{re96,dim03,wa08,li09}, jet model \citep[][]{de15,pri16}, or combination of the two~\citep[][]{yu11,nem14}. Normally, it is believed that the radio emission of M87 dominantly come from jet, while the origin of the millimeter/sub-millimeter and X-ray emission are controversial \citep[either from the ADAF or from the jet,][]{wa08,li09,yu11,pri16}. With ALMA observations, it is found that the radio spectrum becomes much steeper at $\sim 100$ GHz compare to the low-frequency radio band, and the spectrum becomes turnover at $\sim 200$ GHz \citep[][]{pri16}. The similar spectrum of M87 is also found in Sgr A* and M 81 \citep[][]{fal00,yu03,an05,ma08,ma09}. \citet{yu03} proposed that the sub-millimeter bump of Sgr A* can be naturally reproduced by the synchrotron emission from the high-temperature electrons in ADAF. In this work, we get a similar conclusion for M87, which will help us to learn about the underlying accretion process. After constrained by the millimeter bump, we find that the ADAF cannot well reproduce the X-ray emission simultaneously, while the X-ray emission and low-frequency radio emission are better explained by the jet. This conclusion is similar to \citet{yc05,wu07,yua09}, where the X-ray emission should be dominated by the jet, not the ADAF, if the Eddington ratio is less than a critical value. It should be noted that our ADAF-jet model cannot explain the optical-UV emission, which may be contributed by the host galaxy \citep[][]{nem14,de15}, and the multi-waveband flux variations may help to test this issue. We find that different BH spin parameters ($a_{*}=0-0.99$) and magnetic parameters ($\beta=0.5-0.9$) yield an equivalent fit of the SED, but the accretion rate has to be decreased if high BH spin and stronger magnetic field (lower $\beta$) are adopted. The jet velocity also cannot be constrained from our SED modeling, and we find that it will not affect our above conclusion since that it is degenerated with $\dot{m}_{\rm jet}$, where the different jet-velocity parameters will lead to different Doppler factors.  In the ADAF model, the density profile is $\rho\propto r^{-1.5+s}$ and $\rho\propto r^{\sim -1}$ for $s\sim0.4-0.5$, which is quite consistent with that determined by $Chandra$ within the Bondi radius for M87 \citep[][]{russ15}. It was also found that the density profile is quite shallow, $\rho\propto r^{-(0.5-1)}$ in Sgr A* and NGC 3115 \citep[e.g.,][]{wan13,won14}, which is much shallower than that predicted in ``old" ADAF model ($\rho\propto r^{-1.5}$). These results suggest that only a small fraction of the material captured at the Bondi radius reaches the SMBH, which is quite consistent with the recent numerical simulations of the hot flows \citep[e.g.,][and references therein]{yua12}.

   The Faraday RM has been used to constrain the accretion rate in both Sgr A* and M87 \citep[][]{mar06,kuo14}, where they simply assumed a spherical accretion flow surrounding the BH. It may be no problem for Sgr A* due to our LOS is possibly close to the ADAF plane, however, the disk-like ADAF is roughly perpendicular to our LOS in M87 if assuming the jet is perpendicular to the disk (see the cartoon in Figure 1). For this case, the RM cannot be calculated using the same way as Sgr A*. We calculate the RM along the LOS-1 based on the disk-like ADAF (see Figure 1). We find that the RM is around $(0.8-15)\times10^5\rm rad\ m^{-2}$ with different parameters of $a_{*}$ and $\beta$, where the wind parameter $s\simeq0.4-0.5$ has been constrained from SED modeling. Our result is roughly consistent with the observed value of $-(2.1\pm1.8)\times10^5\rm rad\ m^{-2}$ if, in particular, assuming the BH may be fast rotating (e.g., $a_{*}\sim0.9$) in M87 \citep[e.g.,][]{wu07}.  It should be noted that our above conclusion will not change if the RM is calculated along the LOS-1 even in a smaller radius of ADAF (e.g., $2R_{\rm g}<R<10R_{\rm g}$), where the RM values will decrease by a factor of 2. Beside the ADAF model, we also explore the possibility of jet. Due to the jet emission at mm waveband is much larger than that of observation (see Figure 3). Therefore, we only consider the case of jet as an external origin of the RM (e.g., polarized source pass through the jet). The RM should be $<7.5\times10^5\rm rad\ m^{-2}$ if the jet velocity is $>0.99\ c$, where the lower $\dot{m}_{\rm jet}$ is needed for modeling the SED with higher jet velocity. The intrinsic velocity of core jet in M87 is still not known, where the jet may has complex structure, e.g., a fast spine surrounded by a slower layer, and the observed low-velocity is measured from the slower layer \citep[e.g.,][]{gi08,gr09,xie12,na14,wa14,moc16}. Furthermore, the RM will become lower if the magnetic field is strongly dominated by toriodal field in the innermost part of jet or the magnetic field undergo many reversals along LOS. In our model, ADAF model can naturally reproduce the observed RM and we cannot exclude the possibilities of the jet model.  Future constraints on the intrinsic velocity of the spine jet (if the jet is spine-layer structure) will help to further understand this issue.

   Similar to Sgr A* \citep[][]{yu03}, we model the millimeter bump of M87 using a thermal disk component. It should be noted that the millimeter/sub-millimeter bump of both M87 and Sgr A* can also reproduced by the jet component associated with the jet launching region close the BH \citep[so-called ``jet nozzle",][]{fal00,pri16}. In this work, we use a simple jet model with the shape constrained from the observations directly, which do not include such a nozzle. Recently, \citet{li15} calculated the RM of Sgr A* based on the jet nozzle model of \citet{fal00} and found that the predicted RM is two orders of magnitude less than the observed value, which suggest that this model cannot explain the observed RM even it can reproduce the sub-millimeter bump. It is still unknown whether this model can explain the RM of M87 or not, which is beyond our scope.

\section{Summary}
  Using the multi-waveband observational data at a scale of $\sim 0.4$ arcsec, we model the multi-wavelength SED of M87 using a coupled ADAF-jet model, where this model is widely adopted in modeling the SEDs of low-luminosity AGNs. The main results are summarized as follows. \\
   1) We find that the millimeter bump can be naturally reproduced by the synchrotron emission from the thermal electrons in hot accretion flow of the ADAF, where the radio, optical and X-ray emission may still dominantly come from the jet.  \\
   2) The millimeter and submillimeter emission of ADAF mainly come from the inner region of the accretion flow (e.g., $\lesssim 10 R_{\rm g}$), which is roughly consistent with the recent 230\,GHz observations. The density profile of the ADAF ($n_{\rm e} \propto r^{\sim -1}$) is quite consistent with that determined by $Chandra$ within the Bondi radius and the recent numerical experiments. \\
  3) Based on the analysis on the RM, we find that the RM calculated from ADAF with the parameters constrained from the SED modeling is roughly consistent with the measured values.


   \section*{NOTE ADDED IN PRESS}
   After the submission of this manuscript another work has appeared as preprint (Li et al. 2016), which is similar in contents and reaches a very similar conclusion about the accretion and Faraday rotation of M87.
   
   \section*{Acknowledgements}
  This work is supported by the NSFC (grants 11573009, 11133005 and 11303010).


\begin{thebibliography}{}
\bibitem[Abramowicz et al.(1995)]{ab95} Abramowicz, M. A., Chen, X., Kato, S., et al. 1995, \apjl, 438, 37

\bibitem[An et al.(2005)]{an05} An, T., Goss, W. M., Zhao, J.-H., et al. 2005, \apjl, 634, 49

\bibitem[Akiyama et al.(2015)]{ak15}  Akiyama, K., Lu, R.-S., Fish, V. L., et al. 2015, ApJ, 807, 150

\bibitem[Asada \& Nakamura(2012)]{asa12} Asada, K., \& Nakamura, M. 2012, \apj, 745, 28

\bibitem[Asada et al.(2014)]{asa14} Asada, K., Nakamura, M., Doi, A., et al. 2014, ApJL, 781, 2

\bibitem[Broderick \& Loeb(2009)]{br09} Broderick, A. E., \& Loeb, A. 2009, ApJ, 697, 1164

\bibitem[Blakeslee et al.(2009)]{blak09} Blakeslee, J. P., Jord\'{a}n, A., Mei, S., et al. 2009, ApJ, 694, 556

\bibitem[Bower et al.(2003)]{bow03} Bower, G. C., Wright, M. C. H., Falcke, H., et al. 2003, ApJ, 588, 331

\bibitem[Broderick et al.(2011)]{bro11} Broderick, A. E., Loeb, A., Reid, M. J. 2011, ApJ, 735, 57

\bibitem[Cao et al.(2014)]{cao14}  Cao, X.-F., Wu, Q., Dong, A.-J. 2014, ApJ, 788, 52

\bibitem[Coppi \& Blandford(1990)]{cb90} Coppi, P. S., \& Blandford, R. D. 1990, MNRAS, 245, 453

\bibitem[de Jong et al.(2015)]{de15} de Jong, S., Beckmann, V., Soldi, S., et al. 2015, MNRAS, 450, 4333

\bibitem[Dexter et al.(2010)]{dex10} Dexter, J., Agol, E., Fragile, P. C., \& McKinney, J. C. 2010, ApJ, 717, 1092

\bibitem[Dexter et al.(2012)]{dex12} Dexter, J., McKinney, J. C., \& Agol, E. 2012, MNRAS, 421, 1517

\bibitem[Di Matteo et al.(2003)]{dim03} Di Matteo, T., Allen, S. W., Fabian, A. C., et al. 2003, ApJ, 582, 133

\bibitem[Doeleman et al.(2008)]{doe08} Doeleman S. S. et al., 2008, Nature, 455, 78

\bibitem[Doeleman et al.(2009)]{doe09} Doeleman, S., Agol, E., Backer, D., et al. 2009, in ArXiv Astrophysics e-prints, Vol. 2010, astro2010: The Astronomy and Astrophysics Decadal Survey, 68

\bibitem[Doeleman et al.(2012)]{doe12} Doeleman, S. S., Fish, V. L., Schenck, D. E., et al. 2012, Science, 338, 355

\bibitem[Falcke \& Markoff(2000)]{fal00} Falcke, H., \& Markoff, S. 2000, \aap, 362, 113

\bibitem[Fish et al.(2011)]{fish11} Fish, V. L., et al. 2011, ApJ, 727, L36

\bibitem[Gebhardt et al.(2011)]{gebh11} Gebhardt, K., Adams, J., Richstone, D., et al. 2011, ApJ, 729, 119

\bibitem[Giroletti et al.(2008)]{gi08} Giroletti, M., Giovannini, G., Cotton, W. D., et al. 2008, A\&A, 488, 905

\bibitem[Gracia et al.(2009)]{gr09} Gracia, J., Vlahakis, N., Agudo, I., et al. 2009, ApJ, 695, 503

\bibitem[Hada et al.(2016)]{had16} Hada, K., Kino, M., Doi, A., et al. 2016, ApJ, 817, 131

\bibitem[Harris et al.(2006)]{har06} Harris, D. E., Cheung, C. C., Biretta, J. A., et al. 2006, ApJ, 640, 211

\bibitem[Ho(2008)]{ho08} Ho, L. C. 2008, ARA\&A, 46, 475

\bibitem[Huang et al.(2009)]{hua09} Huang, L., Takahashi, R., \& Shen, Z. 2009, ApJ, 706, 960

\bibitem[Huang \& Shcherbakov(2011)]{hua11}  Huang, L., \& Shcherbakov, R. V. 2011, MNRAS, 416, 2574

\bibitem[Ichimaru(1977)]{ic77} Ichimaru, S. 1977, \apj, 214, 840

\bibitem[Jord\'{a}n et al.(2005)]{jord05} Jord\'{a}n, A., C\^{o}t\'{e}, P., Blakeslee, J. P., et al. 2005, ApJ,634, 1002

\bibitem[Junor et al.(1999)]{jun99} Junor, W., Biretta, J. A., \& Livio, M. 1999, Natur, 401, 891

\bibitem[Kuo et al.(2014)]{kuo14} Kuo, C. Y., Asada, K., Rao, R., et al. 2014, ApJ, 783, L33

\bibitem[Kovalev et al.(2007)]{kov07} Kovalev, Y. Y., Lister, M. L., Homan, D. C., et al. 2007, ApJL, 668, L27

\bibitem[Li et al.(2009)]{li09} Li, Y.-R., Yuan, Y.-F., Wang, J.-M., et al. 2009, ApJ, 699, 513

\bibitem[Li, Yuan \&\ Wang(2015)]{li15} Li, Y.-P., Yuan, F., \& Wang, Q.D. 2015, ApJ, 798, 22

\bibitem[Liu \& Wu(2013)]{lw13} Liu, H., \& Wu, Q. 2013, ApJ, 764, 17

\bibitem[Lu et al.(2014)]{lu14} Lu, R.-S., Broderick, A. E., Baron, F., et al. 2014, ApJ, 788, 120

\bibitem[Ly et al.(2007)]{ly07} Ly, C., Walker, R. C., \& Junor, W. 2007, ApJ, 660, 200

\bibitem[Macchetto et al.(1997)]{macc97} Macchetto, F., Marconi, A., Axon, D. J., et al. 1997, ApJ, 489, 579

\bibitem[Marrone et al.(2006)]{mar06} Marrone, D. P., Moran, J. M., Zhao, J.-H., et al. 2006, ApJ, 640, 308

\bibitem[Markoff et al.(2008)]{ma08} Markoff, S., Nowak, M., Young, A., et al. 2008, ApJ, 681, 905

\bibitem[Macquart et al.(2006)]{mac06} Macquart, J.-P., Bower, G. C., Wright, M. C. H., et al. 2006, ApJ, 646, 111

\bibitem[Manmoto(2000)]{man00} Manmoto, T. 2000, \apj, 534, 734

\bibitem[Maitra et al.(2009)]{ma09} Maitra, D., Markoff, S., Brocksopp, C., et al. 2009, MNRAS, 398, 1638

\bibitem[Mocibrodzka et al.(2016)]{moc16} Moscibrodzka, M., Falcke, H., \& Shiokawa, H. 2016, A\&A, 586, 38

\bibitem[Moscibrodzka et al.(2009)]{mos09}  Moscibrodzka, M., Gammie, C. F., Dolence, J. C., et al. 2009, ApJ, 706, 497

\bibitem[Nagai et al.(2014)]{na14} Nagai, H., Haga, T., Giovannini, G., et al. 2014, ApJ, 785, 53

\bibitem[Nagar et al.(2001)]{naga01} Nagar, N.M., Wilson, A.S., \& Falcke, H. 2001, ApJ, 559, L87

\bibitem[Narayan \& Yi(1995)]{ny95} Narayan, R., \& Yi, I. 1995, \apj, 452, 710

\bibitem[Nemmen et al.(2014)]{nem14} Nemmen, R. S., Storchi-Bergmann, T., \& Eracleous, M. 2014, MNRAS, 438, 2804

\bibitem[Piran(1999)]{pi99} Piran, T. 1999, Phys. Rep., 314, 575

\bibitem[Perlman \& Wilson(2005)]{per05} Perlman, E. S., \& Wilson, A. S. 2005, ApJ, 627, 140

\bibitem[Perlman et al.(2001)]{per01} Perlman, E.S., Sparks, W.B., Radomski, J., et al. 2001, ApJ, 561, L51

\bibitem[Prieto et al.(2016)]{pri16} Prieto, M. A., Fernandez-Ontiveros, J. A., Markoff, S., et al. 2016, \mnras, 457, 3801

\bibitem[Quataert \& Gruzinov(2000)]{quat00} Quataert, E., \&\ Gruzinov, A. 2000, ApJ, 545, 842

\bibitem[Reid et al.(1989)]{rei89} Reid, M. J., Biretta, J. A., Junor, W., et al. 1989, ApJ, 336, 112

\bibitem[Ricarte \& Dexter(2015)]{rd15}  Ricarte, A., Dexter, J. 2015, MNRAS, 446, 1973

\bibitem[Reynolds et al.(1996)]{re96} Reynolds, C.S., Di Matteo, T., Fabian, A.C., et al. 1996, MNRAS, 283, L111

\bibitem[Russell et al.(2015)]{russ15} Russell, H. R., Fabian, A. C., McNamara, B. R., et al. 2015, MNRAS, 451, 588

\bibitem[Sharma et al.(2007)]{sh07} Sharma, P., Quataert, E., Hammett, G. W., et al. 2007, \apj, 667, 714

\bibitem[Spada et al.(2001)]{sp01} Spada, M., Ghisellini, G., Lazzati, D., et al. 2001, MNRAS, 325, 1559

\bibitem[Wang et al.(2014)]{wa14}  Wang, J.-Z., Lei, W.-H., Wang, D.-X., et al. 2014, ApJ, 788,32

\bibitem[Wang et al.(2013)]{wan13} Wang, Q. D., Nowak, M. A., Markoff, S. B., et al. 2013, Science, 341, 981

\bibitem[Wang et al.(2008)]{wa08} Wang, J.-M., Li, Y.-R., Wang, J.-C., et al. 2008, ApJ, 676, L109

\bibitem[Wang \& Zhou(2009)]{wz09} Wang, C.-C., \& Zhou, H.-Y. 2009, MNRAS, 395, 301

\bibitem[Walsh et al.(2013)]{wals13} Walsh, J. L., Barth, A. J., Ho, L. C., et al. 2013, ApJ, 770, 86

\bibitem[Whysong \& Antonucci(2004)]{whys04} Whysong, D., \& Antonucci, R. 2004, ApJ, 602, 116

\bibitem[Wong et al.(2014)]{won14}  Wong, K.-W., Irwin, J. A., Shcherbakov, R. V., et al. 2014, \apj, 780, 9

\bibitem[Wu et al.(2013)]{wu13} Wu, Q., Yan, H., Yi, Z. 2013, MNRAS, 436, 1278

\bibitem[Wu et al.(2007)]{wu07} Wu, Q., Yuan,F.,\& Cao, X. 2007, \apj, 669, 96

\bibitem[Wu \& Cao(2006)]{wu06} Wu, Q., \& Cao, X. 2006, PASP, 118, 1098

\bibitem[Xie \& Yuan(2016)]{xy16} Xie, F.-G., \& Yuan, F. 2016, MNRAS, 456, 4377

\bibitem[Xie et al. (2012)]{xie12} Xie, W., Lei, W.-H., Zou, Y.-C., et al. 2012, RAA, 12, 817

\bibitem[Yu et al.(2011)]{yu11} Yu, Z., Yuan, F., \& Ho, Luis C. 2011, ApJ, 726, 87

\bibitem[Yuan \& Narayan(2014)]{yn14} Yuan, F., \& Narayan, R. 2014, ARA\&A, 52, 529

\bibitem[Yuan et al.(2012)]{yua12} Yuan, F., Wu, M., \& Bu, D. 2012, ApJ, 761, 129

\bibitem[Yuan et al.(2009)]{yua09} Yuan F., Yu Z., \& Ho, L. C., 2009, ApJ, 703, 1034

\bibitem[Yuan et al.(2006)]{yu06} Yuan, F., Shen, Z.-Q., \& Huang, L. 2006, \apjl, 642, 45

\bibitem[Yuan \& Cui(2005)]{yc05} Yuan, F., \& Cui, W., 2005, ApJ, 629, 408

\bibitem[Yuan et al.(2003)]{yu03} Yuan, F., Quataert, E., \& Narayan, R. 2003, \apj, 598, 301


\end{thebibliography}
\end{document}